# Diffusion-limited aggregation: A revised mean-field approach

Vladislav A. Bogoyavlenskiy* and Natasha A. Chernova
*Low Temperature Physics Department, Moscow State University, Moscow 119899, Russia*
(Received 16 August 1999; revised manuscript received 30 December 1999)

We propose a revision of the classic mean-field approach of diffusion-limited aggregation (DLA) model originally introduced by Witten and Sander [Phys. Rev. Lett. **47**, 1400 (1981)]. The derived nonlinear mean-field equations providing lattice anisotropy are used to model diffusional growth on square lattice in linear and circular source geometries. The overall cluster shapes obtained from the mean-field calculations are found to satisfy the known scaling behavior experimentally observed for DLA simulations.

## I. INTRODUCTION

The well-known member of the class of stochastic models simulating the Laplacian systems is the diffusion-limited aggregation (DLA) introduced by Witten and Sander [1]. In this theory, self-similar (fractal) ramified patterns grow via irreversible sticking of random ''walkers.'' Despite the simplicity of the DLA rules, the model shows unexpectedly subtle and complex properties and poses a number of theoretical questions such as noise reduction, influence of lattice anisotropy, free-boundary problem, and asymptotic behavior [2–11].

In order to characterize an ensemble-averaged behavior of the DLA model, various mean-field theories (MFT) have been proposed and developed. MFT is a set of evolution equations illustrating a general continuous formulation for the time development of growing clusters. Constructing of a MFT raises the following main problem: how can be realized the correct transition from discrete units to finite walker and cluster distributions? A first attempt to establish a mean-field approach for the DLA model goes back to the pioneering work by Witten and Sander [1]. Performing a continuous formulation of walker ($u$) and cluster ($\rho$) mean densities, they proposed the following equations:

$$\frac{\partial u}{\partial t} = \nabla^2 u - \frac{\partial \rho}{\partial t}, \qquad (1)$$

$$\frac{\partial \rho}{\partial t} = u(\rho + a^2 \nabla^2 \rho), \qquad (2)$$

where $a$ is the lattice parameter. In this set of relations, Eq. (1) represents the conservation of mass in a diffusive system, and Eq. (2) accounts for the growing rule of the cluster field.

Unfortunately, the Witten-Sander theory cannot model a stable front of clusters due to the crucial instability of Eq. (2): a small perturbation in the $\rho$ field, in the presence of $u$, will grow exponentially [12]. This unstable behavior is caused by the lack of a threshold in this continuous approach. However, the discrete DLA model has an implicit threshold: growth at a site is disallowed unless a nearest-neighbor site is fully occupied. In order to remedy the insufficiencies of the classic model, Brener *et al.* proposed some modification of Eq. (2) which consists in replacing $\rho$ by $\rho^\gamma$ [13]:

$$\frac{\partial \rho}{\partial t} = u(\rho^\gamma + a^2 \nabla^2 \rho). \qquad (3)$$

Taking $\gamma$ greater than 1 is a way to introduce a cutoff in the growth rate at small cluster density. As argued in Ref. [13], any function $F(\rho)$ that vanishes faster than linearly as $\rho \to 0$ can be also used instead of $\rho^\gamma$, e.g., $F(\rho) = \rho \Theta(\rho - A)$ (where $\Theta$ is the Heaviside function). The substitution mimics the fact that in the DLA model, the cluster growth cannot occur with an infinitesimal fluctuation of $\rho$ field. This $\gamma$ model demonstrates a steady-state growth in channel [13] and sector [14] geometries, and can be also used for simulations of convex-concave morphological transitions in diffusive systems [15].

Nevertheless, the $\gamma$-MFT has two important intrinsic problems. First, the theory motivates the question of how to obtain a growth threshold in a more fundamental way. In the $\gamma$ model, there is no global revision of the Witten-Sander approach, the authors just proposed a mathematical substitution $\rho^\gamma \leftrightarrow \rho$ [it should be emphasized that the phenomenological parameter $\gamma \in (2 \cdots 10)$ does not have a clear physical explanation]. In addition, the phenomenological anisotropy introduced by $[\partial^2 \rho / \partial x^2 + b(\partial^2 \rho / \partial y^2)]$- or $[\partial^4 \rho / \partial x^4 + \partial^4 \rho / \partial y^4]$-like terms in Eq. (3) also raises questions about its derivation [13–15]. The second problem seems to be more serious. The comparison between the $\gamma$-MFT predictions and the occupancy probability distributions computed from DLA simulations yields satisfactory results as long as the ensemble averaging is performed on small-size DLA clusters. Some severe discrepancies arise when one proceeds to large-size DLA simulations; as pointed out by Arneodo *et al.* [16], the $\gamma$ model fails to reproduce the spreading of active front zone of the clusters grown in wide channels or in divergent sector cells. The width of active front $\Delta$ computed from the mean-field equations does not display any time dependence, a result that is in contradiction with the scaling behavior known for DLA clusters

$$\Delta \sim X_F^{1/2}, \qquad (4)$$

---

where $X_F$ is the front position. This is, without any doubt, one of the main weaknesses of the $\gamma$-MFT.

In order to understand the inadequacy of the theories discussed, one has to come back to the original work by Witten and Sander [1], and to revise one of the main ingredients of the classic approach. There have been several attempts to construct a MFT with the use of some alternative assumptions, e.g., of the cluster nonpenetrability (deterministic) [17] or of the generic noise effect (stochastic) [18]. In the present work, our goal is to revise the Witten-Sander MFT in terms of the Boltzmann theory of irreversible processes. The paper is organized as follows. In Sec. II, a revised mean-field approach of the DLA model is introduced. The subject of Sec. III is the comparison between the mean-field predictions and results of an ensemble averaging of DLA clusters. Finally, in Sec. IV a summary of the work is given.

## II. MEAN-FIELD APPROACH

### A. General theory

According to the general Boltzmann theory of irreversible processes, the DLA model can be considered as a two-particle interaction between walker $u(\mathbf{r},t)$ and cluster $\rho(\mathbf{r},t)$ fields. In order to describe the process of aggregation, let us write the interaction intensity $St(\mathbf{r},t) = \partial \rho(\mathbf{r},t)/\partial t$ (the Boltzmann integral) as

$$St(\mathbf{r},t) = \int_{\mathbf{r}+\mathbf{e}\in I} u(\mathbf{r},t)\rho(\mathbf{r}+\mathbf{e},t) W_{\text{int}}(\mathbf{r},\mathbf{e},t) dI. \quad (5)$$

Here $u(\mathbf{r},t)$ and $\rho(\mathbf{r},t)$ are considered as the distribution functions

$$0 \le u(\mathbf{r},t) \le 1, \quad 0 \le \rho(\mathbf{r},t) \le 1. \quad (6)$$

In Eq. (5), the function $W_{\text{int}}(\mathbf{r},\mathbf{e},t)$ represents the probability of the successful interaction (i.e., leading to the aggregation) between walker and cluster fields; the integrating is performed inside the sphere of interaction $\mathbf{r}+\mathbf{e}\in I$. This formulation can be explained as follows. If there is a two-particle interaction where the first particle is a walker ''unit'' $u(\mathbf{r},t)$ and the second one is a cluster ''unit'' $\rho(\mathbf{r}+\mathbf{e},t)$, then the integration means that the walker unit interacts with all possible neighboring cluster units.

The issue of this theory is to establish the relationship between $W_{\text{int}}(\mathbf{r},\mathbf{e},t)$ and the interacting fields. In order to understand the nature of the aggregation process, let us focus on a particular walker unit $u(\mathbf{r}_0)$ which interacts with a cluster unit $\rho(\mathbf{r}_0+\mathbf{e}_0)$. In the classic Witten-Sander mean-field approach, the interaction probability is considered to be number one, $W_{\text{int}}(\mathbf{r}_0,\mathbf{e}_0) \equiv 1$, i.e., even for an infinitesimal value of the cluster density $\rho(\mathbf{r}_0+\mathbf{e}_0)$ each interaction with the walker unit $u(\mathbf{r}_0)$ leads to the aggregation. This assumption seems to be the most questionable. It is probable that a more realistic hypothesis is the linear connection between $W_{\text{int}}(\mathbf{r}_0,\mathbf{e}_0)$ and $\rho(\mathbf{r}_0+\mathbf{e}_0)$, i.e., the probability of the successful interaction is proportional to the cluster density. Due to the normalization condition [Eq. (6)], we propose the following main relation of the revised mean-field approach:

$$W_{\text{int}}(\mathbf{r},\mathbf{e},t) = \rho(\mathbf{r}+\mathbf{e},t). \quad (7)$$

In this way, we introduce a threshold for the cluster growth. As a result, Eq. (5) transforms to

$$\frac{\partial \rho(\mathbf{r},t)}{\partial t} = u(\mathbf{r},t) \int_{\mathbf{r}+\mathbf{e}\in I} \rho^2(\mathbf{r}+\mathbf{e},t) dI. \quad (8)$$

### B. On-lattice model

The proposed relation (8) gives a general kinetics of the mean-field model. It contains two variable parameters: a sphere of interactions $I$ and a vector set $\mathbf{e}$. These parameters are determined by conditions of neighborhood (e.g., off-lattice, on-lattice) of the DLA. In the present paper, we restrict our study by a DLA model with the following properties: (i) the aggregation takes place on a lattice, i.e., the integrating can be replaced by a finite summation and (ii) the vector set $\mathbf{e}$ has a center of symmetry, i.e., $\Sigma_i \mathbf{e}_i = 0$. Then we can rewrite Eq. (8) as

$$\frac{\partial \rho(\mathbf{r},t)}{\partial t} = u(\mathbf{r},t) \sum_i \rho^2(\mathbf{r}+\mathbf{e}_i,t). \quad (9)$$

Here $\mathbf{e}_i$ are the vectors to adjacent sites and $i$ runs over the number of neighbors.

In order to obtain the continuous representation of Eq. (9), let us use the formula of the Taylor decomposition

$$\rho(\mathbf{r}+\mathbf{e}_i,t) = \rho(\mathbf{r},t) + \mathbf{e}_i \nabla \rho(\mathbf{r},t) + \frac{1}{2}\mathbf{e}_i \nabla [\mathbf{e}_i \nabla \rho(\mathbf{r},t)]. \quad (10)$$

The squared Eq. (10) follows from the expression

$$\rho^2(\mathbf{r}+\mathbf{e}_i,t) = \rho^2 + \rho[2\mathbf{e}_i\nabla\rho + \mathbf{e}_i\nabla(\mathbf{e}_i\nabla\rho)] + [\mathbf{e}_i\nabla\rho]^2$$
$$+ [\mathbf{e}_i\nabla\rho][\mathbf{e}_i\nabla(\mathbf{e}_i\nabla\rho)] + \frac{1}{4}[\mathbf{e}_i\nabla(\mathbf{e}_i\nabla\rho)]^2, \quad (11)$$

where we write $\rho$ instead of $\rho(\mathbf{r},t)$. After the $i$ summation, the terms with odd powers of $\mathbf{e}_i$ are rejected (due to the condition of lattice symmetry $\Sigma_i \mathbf{e}_i = 0$), and we obtain

$$\sum_i \rho^2(\mathbf{r}+\mathbf{e}_i,t) = \sum_i \left\{ \rho^2 + \rho[\mathbf{e}_i\nabla(\mathbf{e}_i\nabla\rho)] + [\mathbf{e}_i\nabla\rho]^2 \right.$$
$$\left. + \frac{1}{4}[\mathbf{e}_i\nabla(\mathbf{e}_i\nabla\rho)]^2 \right\}. \quad (12)$$

As a result, Eq. (9) transforms to the following differential law for the time evolution of the cluster field:

$$\frac{\partial \rho}{\partial t} = u \sum_i \left\{ \rho^2 + \rho[\mathbf{e}_i\nabla(\mathbf{e}_i\nabla\rho)] + [\mathbf{e}_i\nabla\rho]^2 \right.$$
$$\left. + \frac{1}{4}[\mathbf{e}_i\nabla(\mathbf{e}_i\nabla\rho)]^2 \right\}. \quad (13)$$

One can notice two main properties of the mean-field relation derived. First, Eq. (13) demonstrates the linear stability to infinitesimal fluctuations; a small perturbation $\delta\rho$ in the $\rho(\mathbf{r},t)$ field will vanish as

$$\frac{\partial(\delta\rho)}{\partial t} \sim (\delta\rho)^2. \tag{14}$$

Second, Eq. (13) provides lattice anisotropy due to the anisotropic features of the term $[\mathbf{e}_i \nabla(\mathbf{e}_i \nabla \rho)]^2$.

## III. NUMERICAL SIMULATIONS

The mean-field equation (13) derived in the previous section represents a general relation that can be used for different dimensions (1D, 2D, 3D, ...), source geometries (linear, circular, spherical, ...), and lattices (square, hexagonal, cubic, ...). In this work, we present a study of the mean-field model on square lattice. Assuming only nearest-neighbor interactions, we can rewrite Eq. (13) as

$$\frac{\partial \rho}{\partial t} = u \left\{ \rho^2 + a^2 \left[ \rho \left( \frac{\partial^2 \rho}{\partial x^2} + \frac{\partial^2 \rho}{\partial y^2} \right) + \left( \frac{\partial \rho}{\partial x} \right)^2 + \left( \frac{\partial \rho}{\partial y} \right)^2 \right] + \frac{a^4}{4} \left[ \left( \frac{\partial^2 \rho}{\partial x^2} \right)^2 + \left( \frac{\partial^2 \rho}{\partial y^2} \right)^2 \right] \right\}, \tag{15}$$

where $a \equiv (\Sigma_i \mathbf{e}_i^2)^{1/2}$ is the lattice spacing. To obtain the complete set, Eq. (15) should be coupled with the walker diffusion equation (1) written for the case of square lattice as

$$\frac{\partial u}{\partial t} = \left( \frac{\partial^2 u}{\partial x^2} + \frac{\partial^2 u}{\partial y^2} \right) - \frac{\partial \rho}{\partial t}. \tag{16}$$

### A. Linear source geometry

Let us assume that motion and aggregation of walkers take place inside a channel of width $W$. As argued in Ref. [13], the $\gamma$-MFT gives the distribution of cluster field in the asymptotic case $x \to \infty$, when the behavior of mean-field equations is fully determined by the lateral boundaries (i.e., width of the active front zone $\Delta \to W$). In this case, the theory satisfactorily describes the Saffman-Taylor patterns [13]. However, the $\gamma$ model fails to predict the cluster shape when the boundaries do not affect the cluster field, i.e., in the case $\Delta < W/2$ [16]. In order to examine the mean-field equations introduced, we present the study of this wide-channel problem.

#### 1. Mean-field predictions

Formulating the restrictions on the lateral boundaries of the channel, we impose the Neuman condition for the $u$ field and the Dirichlet condition for the $\rho$ field:

$$\frac{\partial u}{\partial y} = 0 \quad \text{as} \quad y = \pm \frac{W}{2}, \tag{17}$$

$$\rho = 0 \quad \text{as} \quad y = \pm \frac{W}{2}. \tag{18}$$

At infinity, we fix the flux of the walkers

$$\frac{\partial u}{\partial x} = \varphi \quad \text{as} \quad x \to \infty. \tag{19}$$

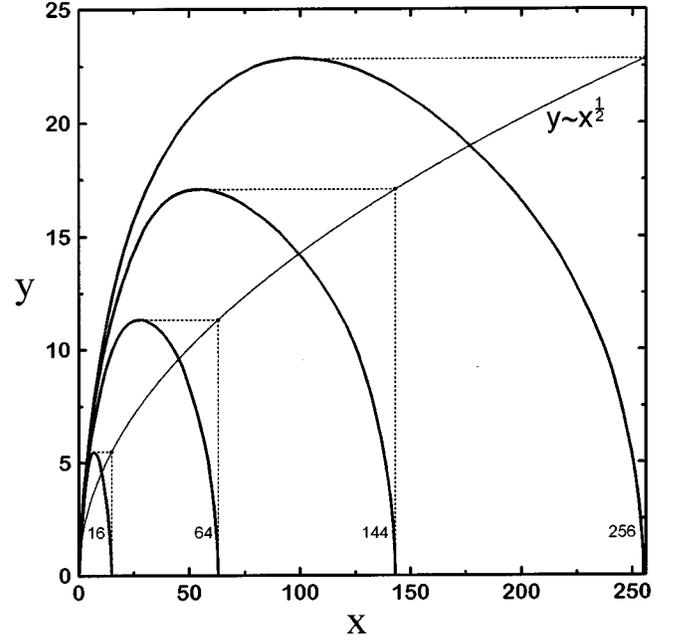

FIG. 1. Mean-field calculations of overall cluster shape performed inside channel of width $W=128$ on square lattice ($a=1$) in linear source geometry. Thick curves represent contours $\rho(x,y \geq 0) = 0.01$ of cluster field distribution at different stages of growth; numbers at the curves represent positions of active front zone $X_F$. Thin auxiliary curve is function $y \sim x^{1/2}$ illustrating by dotted lines the scaling behavior of front width $\Delta \sim X_F^{1/2}$.

Let us consider the cluster growth to settle in the center of the channel from the beginning:

$$\rho(0,0) = 1. \tag{20}$$

For this initial condition, the scaling behavior $\Delta \sim X_F^{1/2}$ [Eq. (4)] is known from DLA simulations. In order to prove the validity of the mean-field equations derived, we have solved Eqs. (15)–(20) numerically. In Fig. 1, we present the overall contour plots of cluster field distribution $\rho(x,y \geq 0) = 0.01$ calculated for the channel of width $W=128$ at different stages of growth. The figure demonstrates the evolution of cluster shape at the positions of the active front zone $X_F = 16, 64, 144$, and 256. For each stage of the cluster growth, one can observe the classic Ivantsov law [19] in the tip regions of the curves. As a result, the cluster width $\Delta$ exactly resembles the scaling $\Delta \sim X_F^{1/2}$ illustrated by the auxiliary curve $y \sim x^{1/2}$. It should be noted that similar contour plots were obtained by Kassner and Brener [20] when they constructed the noiseless DLA model from the macroscopic principles of the cluster tip formation.

In Fig. 2, we present the spatial cluster distribution in the case of $X_F = 256$. The figure shows the contour plots of the cluster field $\rho(x,y)$ [Fig. 2(a)], the longitudinal profile of $\rho(x,y)$ in the section $y=0$ [Fig. 2(b)], and the transverse profile of $\bar{\rho}(x,y) \equiv \rho(x,y)/\rho(100,0)$ in the section $x=100$ [Fig. 2(c)]. The obtained contour plots are characterized by a fingerlike shape where the tip length slightly exceeds the base one, so the maximal cluster width $\Delta_{\max} \approx 46$ is observed in a neighborhood of the section $x=100$. In this section, the transverse cluster profile [Fig. 2(c)] demonstrates a convex-up behavior; the maximum of the density is related to the cluster axe. As shown by Fig. 2(b), there are three distinct regions in the longitudinal profile of the cluster field. First there is some initial transient regime $x \in (0 \cdots 50)$ where one progressively loses the influence of initial conditions to the benefit of the growth. In this region, the cluster density decreases from 1 to a constant value $\rho \approx 0.6$. Then, there is a region of a stable growth $x \in (50 \cdots 240)$ where the cluster field changes insignificantly. In the third region $x \in (240 \cdots 256)$, there is a rapid falloff of the density.

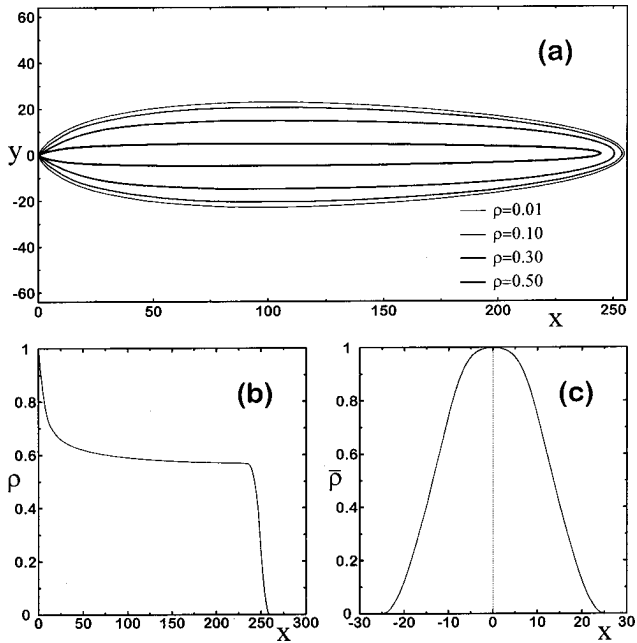

FIG. 2. Mean-field calculations of cluster field distribution $\rho(x,y)$ performed inside channel of width $W=128$ on square lattice ($a=1$) in linear source geometry. (a) Contour plots of $\rho(x,y)$ where the levels are $\rho(x,y) = 0.01$, 0.1, 0.3, and 0.5 from the outer to the inner. (b) Longitudinal profile of $\rho(x,y)$ in the section $y=0$. (c) Transverse profile of $\bar{\rho}(x,y) = \rho(x,y)/\rho(100,0)$ in the section $x=100$.

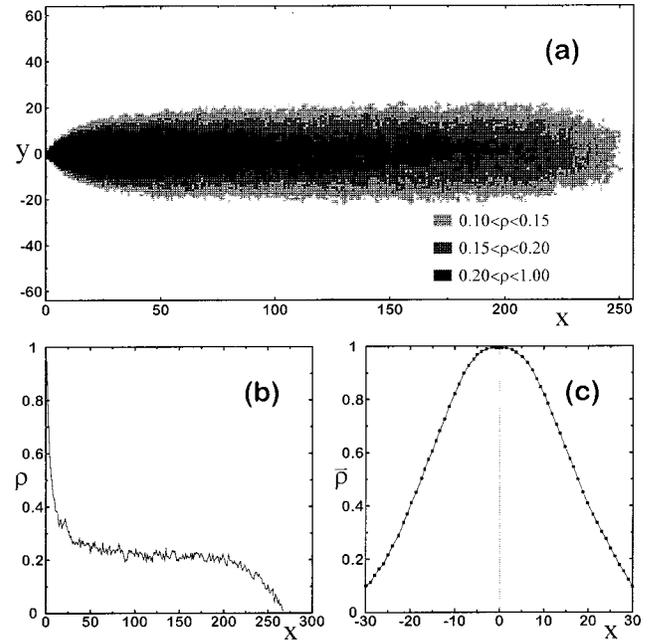

FIG. 3. Statistical analysis of $N=1000$ DLA clusters containing $M=2\times 10^3$ particles grown inside channel of width $W=128$ on square lattice ($a=1$) in linear source geometry. (a) Mean cluster occupancy $\rho(x,y)$ in field-plot representation where the fields are $0.1 \leq \rho(x,y) < 0.15$, $0.15 \leq \rho(x,y) < 0.2$, and $0.2 \leq \rho(x,y) < 1$ from the outer to the inner. (b) Longitudinal profile of $\rho(x,y)$ in the section $y=0$. (c) Transverse profile of $\bar{\rho}(x,y) = \rho(x,y)/\rho(100,0)$ averaged over 20 sections $x \in (90 \cdots 110)$.

*2. Ensemble averaging*

In order to compare the theoretical mean-field predictions with results of a statistical analysis of DLA clusters, one has to measure the mean occupancy distribution $\rho(x,y)$ obtained from an ensemble averaging over a given number of clusters $N$. For Monte Carlo (MC) simulations, we use the classic Witten-Sander algorithm [1]. The walkers are released from a linear source outside the cluster; when a walker becomes adjacent to the cluster, the walker site is considered to be occupied. In a given channel, we grow $N$ aggregates with the same total number $M$ of particles. We then count for each site how many times it has been occupied by a particle of a cluster. The mean occupancy $\rho(x,y)$ is obtained by dividing this number by the total number $N$ of realizations.

In Fig. 3, we present the analysis of the ensemble averaging of $2 \times 10^3$-particle DLA clusters simulated in the channel of width $W=128$. In order to decrease noise errors, the results are averaged over 1000 clusters. The figure demonstrates a two-dimensional representation of the cluster distribution $\rho(x,y)$ [Fig. 3(a)], the longitudinal profile of $\rho(x,y)$ in the section $y=0$ [Fig. 3(b)], and the transverse profile of

$\bar{\rho}(x,y) \equiv \rho(x,y)/\rho(100,0)$ in a neighborhood of the section $x=100$ [Fig. 3(c)]. The obtained cluster distribution $\rho(x,y)$ and its profiles qualitatively resemble the results of the theoretical predictions [Fig. 2]. To resume the difference between the figures, the theoretical distribution $\rho(x,y)$ is characterized by more sharp behavior than the statistical one. The most discrepancy is related to the tip region of the longitudinal cluster profile [Figs. 2(b) and 3(b)] and to the tails of the transverse cluster profile [Figs. 2(c) and 3(c)] where the observed dispersion of the statistical results drastically hampers the detailed comparison.

To reduce the effect of noise, one of the possible ways is to increase the number of realizations $N$. However, the dependence of the noise error $\delta_N$ on the number $N$ is rather weak:

$$\delta_N \sim \frac{1}{\sqrt{N}}. \quad (21)$$

To decrease the noise amplitude by 10 times, one has to increase the number of clusters by 100 times. This way seems to be impractical because of computational time limitation. In order to decrease the noise influence in a more efficient way, we combine the ensemble averaging with the noise-reducing algorithm introduced by Tang [3]. Rather than take a single walk as an independent contribution to the cluster, a multiple registration for every interfacial site is considered. The site can be occupied only when it has been registered $N_R$ times; this scheme results in the reduction of spatial dispersion without the increase of total simulation time [5–7].

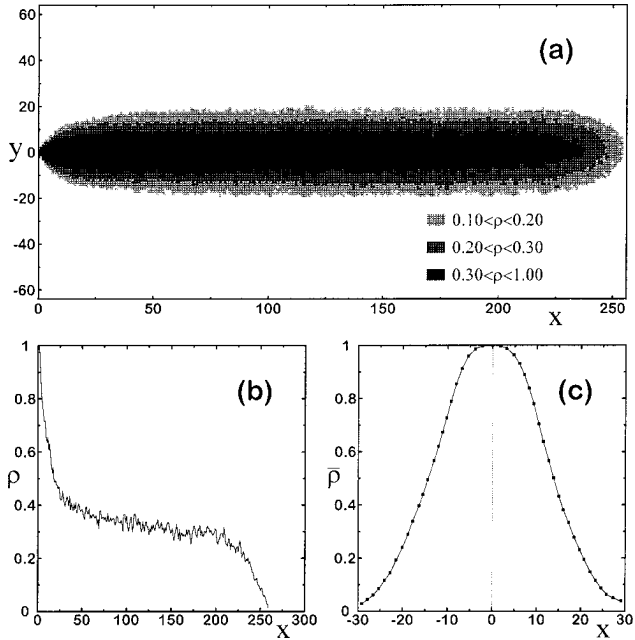

FIG. 4. Statistical analysis of $N=250$ DLA clusters (obtained by Tang's averaging scheme with number of registrations $N_R=4$) containing $M=2\times 10^3$ particles grown inside channel of width $W=128$ on square lattice ($a=1$) in linear source geometry. (a) Mean cluster occupancy $\rho(x,y)$ in field-plot representation where the fields are $0.1\leq\rho(x,y)<0.2$, $0.2\leq\rho(x,y)<0.3$, and $0.3\leq\rho(x,y)<1$ from the outer to the inner. (b) Longitudinal profile of $\rho(x,y)$ in the section $y=0$. (c) Transverse profile of $\bar\rho(x,y)=\rho(x,y)/\rho(100,0)$ averaged over 20 sections $x\in(90\cdots 110)$.

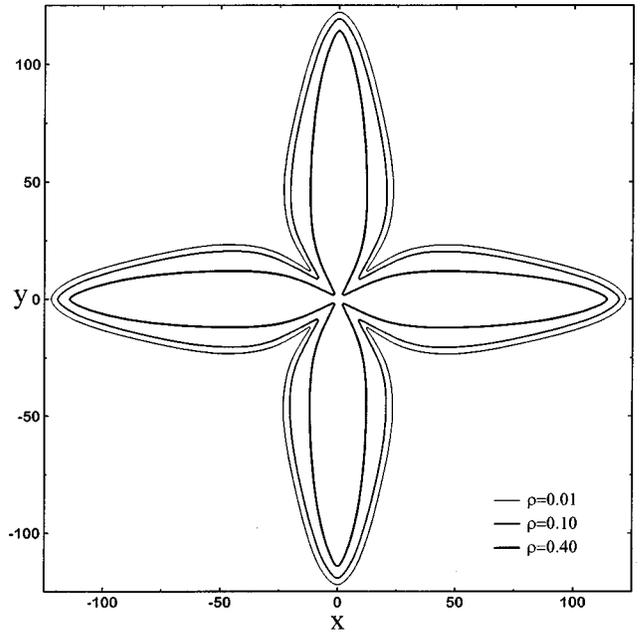

FIG. 5. Contour plots of cluster field distribution $\rho(x,y)$ obtained from mean-field calculations on square lattice ($a=1$) in circular source geometry. Contour levels are $\rho(x,y)=0.01$, 0.1, and 0.4 from the outer to the inner.

The application of Tang's noise-reducing algorithm to the ensemble averaging is summarized by Fig. 4, which illustrates the mean cluster occupancy $\rho(x,y)$ [Fig. 4(a)] and also its longitudinal [Fig. 4(b)] and transverse [Fig. 4(c)] profiles in the same way as Fig. 3. Even for a small number of registrations $N_R=4$ the obtained dispersion of the cluster field significantly decreases in comparison to one observed without Tang's scheme [Fig. 3]. The mean cluster occupancy is characterized by the same fingerlike shape [Figs. 2(a) and 3(a)], and its longitudinal [Fig. 4(b)] and transverse [Fig. 4(c)] profiles seem to be described by the theoretical curves [Figs. 2(b) and 2(c)] in a more precise way, especially in the tip of the cluster and in the tails of the transverse profile.

### B. Circular source geometry

#### 1. Mean-field predictions

Let us assume that motion and aggregation of the growth units take place inside a circle. In this case, we have only one boundary condition at infinity where the flux of the walkers is fixed:

$$\frac{\partial u}{\partial r} = \varphi \quad \text{as} \quad r\to\infty. \tag{22}$$

Here $r\equiv\sqrt{x^2+y^2}$ is the distance from point $(x,y)$ to the origin. As the initial condition, we consider the cluster nucleus located at the origin, $\rho(0,0)=1$ [Eq. (20)].

The results of the mean-field calculations are summarized by Fig. 5, which shows the contour plots of theoretical distribution $\rho(x,y)$. Each plot can be described as four symmetric fingerlike branches that grow in directions $\langle 10\rangle$, $\langle\bar 10\rangle$, $\langle 01\rangle$, and $\langle 0\bar 1\rangle$ due to the square lattice symmetry. The branches qualitatively resemble the shapes obtained in the linear source geometry [Fig. 2(a)]; only the width to length ratios differ. In the region of active front zone, the cluster field is also observed to satisfy the scaling behavior $\Delta\sim X_F^{1/2}$ [Eq. (4)].

#### 2. Ensemble averaging

For the statistical analysis, we average the cluster ensemble simulated by the classic Witten-Sander DLA algorithm [1] and by Tang's noise-reducing scheme [3] where the walkers are released from a circular source outside the cluster. In Fig. 6, we present the mean cluster occupancy $\rho(x,y)$ of $4\times 10^3$-particle DLA clusters averaged over 1000 simulations. The simple ensemble averaging results in a significant noise dispersion of data that hampers the following analysis of the cluster distribution. This dispersion drastically exceeds one observed in linear source geometry (Fig. 3); one can resolve only an anisotropic behavior of the cluster field.

The mean cluster occupancy $\rho(x,y)$ of $4\times 10^3$-particle DLA clusters shown in Fig. 7 is obtained from MC simulations where the averaging over 250 clusters is combined with the additional Tang's noise-reducing scheme (the number of registrations $N_R=4$). In contrast to Fig. 6, the noise reduction gives an opportunity to characterize the cluster field in a more detailed way. As seen in the figure, the obtained cluster distribution $\rho(x,y)$ satisfactorily resembles the theoretical prediction (Fig. 5).

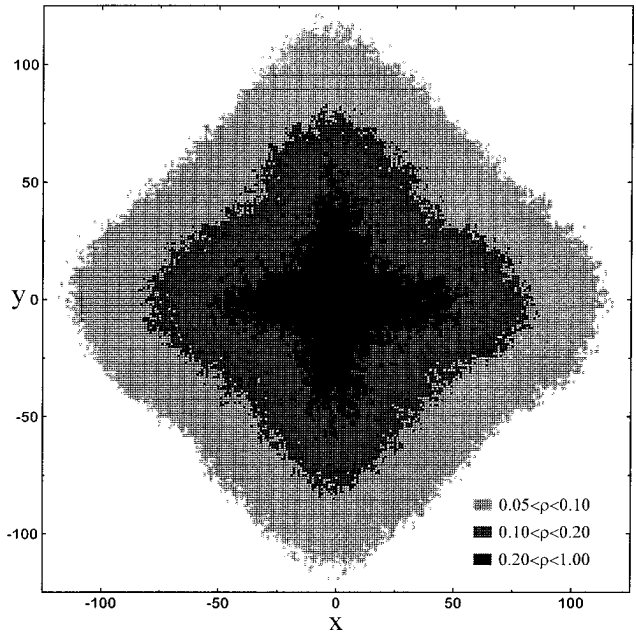

FIG. 6. Field-plot representation of mean cluster occupancy $\rho(x,y)$ of $N=1000$ DLA clusters containing $M=4\times 10^3$ particles grown on square lattice ($a=1$) in circular source geometry. The fields are $0.05 \leq \rho < 0.1$, $0.1 \leq \rho < 0.2$, and $0.2 \leq \rho < 1$ from the outer to the inner.

### 3. Stochastic behavior of diffusive field

In conclusion of the paper, let us investigate the mean-field approach introduced [Eq. (9)] in the case of stochastic behavior of the diffusive field $u$. This problem is consistent with the following MC algorithm. Initially, a nucleus of the cluster is located at the origin, $\rho(0,0)=1$. Each walker modeled by a simple random motion is considered to transfer a constant density $u_0 \ll 1$ (in our simulations, $u_0 = \frac{1}{256}$). When a walker becomes adjacent to the cluster, the attachment probability $P$ is calculated by the formula

$$P = \rho^2(x+1,y) + \rho^2(x-1,y) + \rho^2(x,y+1) + \rho^2(x,y-1). \quad (23)$$

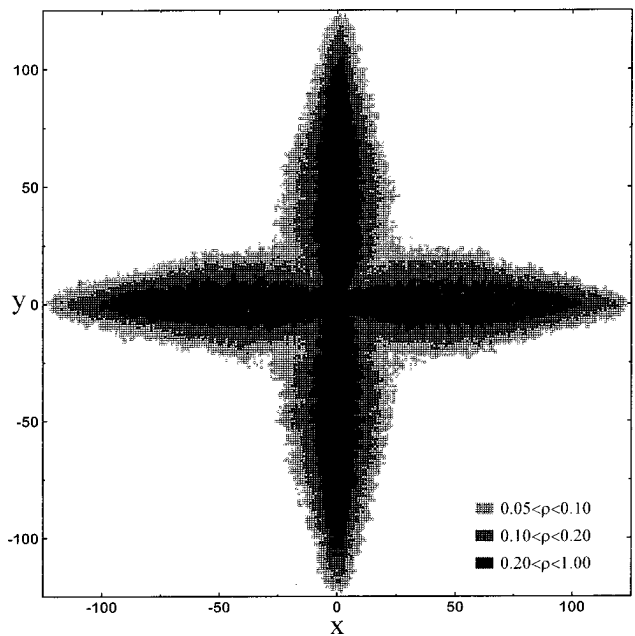

FIG. 7. Field-plot representation of mean cluster occupancy $\rho(x,y)$ of $N=250$ DLA clusters (obtained by Tang's averaging scheme with number of registrations $N_R=4$) containing $M=4\times 10^3$ particles grown on square lattice ($a=1$) in circular source geometry. The fields are $0.05 \leq \rho < 0.1$, $0.1 \leq \rho < 0.2$, and $0.2 \leq \rho < 1$ from the outer to the inner.

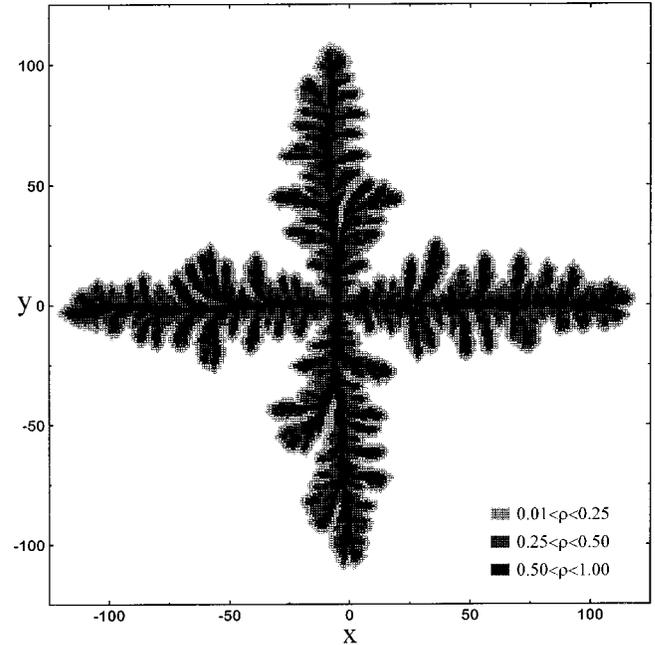

FIG. 8. Field-plot representation of cluster field distribution $\rho(x,y)$ obtained in case of stochastic walker behavior on square lattice ($a=1$) in circular source geometry. The fields are $0.01 \leq \rho < 0.25$, $0.25 \leq \rho < 0.5$, and $0.5 \leq \rho < 1$ from the outer to the inner.

Then a random number $0 < R < 1$ is simulated. If $R < P$, the walker transforms into the cluster site and advances the cluster density $\rho(x,y)$ by the value of $u_0$. In the opposite case, $R > P$, that walker continues its random motion until it aggregates somewhere on the cluster. As successive walkers repeat this process, the cluster density field $\rho(x,y)$ is modified. The simulation is continued until the cluster reaches a specified size.

The results of MC simulations are summarized by Fig. 8, which illustrates the cluster distribution in two-dimensional representation. The features of the obtained cluster field significantly depend on the length scale. On the one hand, the overall cluster shape satisfies the mean-field prediction shown in Fig. 5. The pattern can be described as a compact fingerlike structure with fourfold symmetry; near the growth front, the cluster density rapidly increases from 0 to an average value $\rho \approx 0.25$. One the other hand, the cluster distribution also demonstrates a ramified behavior; within the pattern, a fractal DLA-like ''skeleton,'' $\rho(x,y) \geq 0.5$, can be resolved. To resume, the observed cluster evolution includes a superposition of deterministic and random processes. The deterministic part is governed by the mean-field equations introduced; the random one follows the stochastic DLA behavior. As a consequence, we observe the effects of side-branching and tip splitting that make the obtained pattern very similar to ones described in most diffusive systems [21,22].

## IV. SUMMARY

(1) We have revised the Witten-Sander mean-field approach of the DLA model in terms of the Boltzmann theory of irreversible processes. Instead of the classic linear connection between the DLA intensity and the neighboring cluster densities, we propose the squared law followed from the phenomenological assumption. The derived mean-field equations demonstrate both the front stability to infinitesimal fluctuations and the anisotropic behavior caused by the nonlinearity of the approach introduced.

(2) We have examined the proposed mean-field equations to satisfy the scaling behavior for width of the active cluster zone experimentally known from DLA simulations.

(3) We have studied the ensemble averaging of DLA clusters grown on square lattice in linear and circular geometries of source. The comparison between the mean-field predictions and the ensemble averaging gives a qualitative resemblance of the data. The discrepancy significantly decreases when one proceeds to Tang's noise-reducing DLA algorithm.

(4) We have investigated the influence of stochastic walker behavior on cluster field distribution. The overall cluster shape is observed to satisfy the mean-field prediction calculated for deterministic walker motion. The local cluster distribution is characterized by fractal properties known for DLA structures; this results in the sidebranching and tip splitting.